# THE FALLACY OF THE CLOSEST ANTENNA:

## TOWARDS AN ADEQUATE VIEW OF DEVICE LOCATION IN THE MOBILE NETWORK

Aleksey Ogulenko[1], Itzhak Benenson[1]*, Marina Toger[2], John Östh[2,3], and Alexey Siretskiy[2]

ogulenko.a.p@gmail.com, marina.toger@kultgeog.uu.se,

john.osth@kulteog.uu.se, bennya@tauex.tau.ac.il

[1]Department of Geography and Human Environment,
Porter School of the Environment and Earth Sciences, Tel Aviv University

[2]Department of Social and Economic Geography, Uppsala University

[3]Department of Civil Engineering and Energy Technology, Oslo Metropolitan University

*Corresponding author

## Abstract

The partition of the Mobile Phone Network (MPN) service area into the cell towers' Voronoi polygons (VP) may serve as a coordinate system for representing the location of the mobile phone devices. This view is shared by numerous papers that exploit mobile phone data for studying human spatial mobility. We investigate the credibility of this view by comparing volunteers' locational data of two kinds: (1) Cell towers' that served volunteers' connections and (2) The GPS tracks of the users at the time of connection. In more than 60% of connections, user's mobile device was found outside the VP of the cell tower that served for the connection. We demonstrate that the area of possible device's location is many times larger than the area of the cell tower's VP. To comprise 90% of the possible locations of the devices that may be connected to the cell tower one has to consider the tower's VP together with the two rings of the VPs adjacent to the tower's VPs. An additional, third, ring of the adjacent VPs is necessary to comprise 95% of possible locations of the devices that can be connected to a cell tower.

The revealed location uncertainty is in the nature of the MPN structure and service and entail essential overlap between the cell towers' service areas. We discuss the far-reaching consequences of this uncertainty in regards to the estimating of locational privacy and urban mobility. Our results undermine today's dominant opinion that an adversary, who obtains the access to the database of the Call Detail Records maintained by the MPN operator, can identify a mobile device without knowing its number based on a very short sequence of time-stamped field observations of the user's connection.

## 1. Introduction

The Voronoi partition of the Mobile Phone Network (MPN) service area is an iconic starting point of numerous papers that exploit the mobile phone data for studying human mobility (Berlingerio et al. 2013; Shin et al, 2015, Pinelli et al. 2016; Markovic et al. 2017, Fan et al, 2018, Huang et al, 2018, Wang et al, 2018, Bachir et al, 2019). The source of this view is in the Bell Systems' pioneer developments that started with the Ring's memorandum of 1949 (Ring, 1947) and concluded in MacDonald's paper of 1979 who consider an abstract hexagonal partition of the service space in order to present the idea of the mobile telephony. The earliest publication we are aware of that makes a step from the hexagonal partition to Voronoi polygons, also comes from the telephone industry (INRIA Sophia-Antipolis and France TELECOM team) and deals with the MPN service planning (Baccelli et al, 1997).

MPN developers use hexagonal/Voronoi coverage for presenting their technical ideas. However, researchers that analyze mobile phone data made an additional step and, explicitly or implicitly, assumed that the mobile device is served by the closest MPN antenna and at the moment of connection, and, therefore, the user is located within the Voronoi polygon of a cell tower on which this antenna is mounted. Importantly, in many countries, the map of cell towers is open to the public (see [www.cellmapper.net](www.cellmapper.net)) and can be easily turned into the tower-based VP coverage. That makes tower-based VPs coverage an open and easily available coordinate system that can be used for representing users' location and trajectories. The precision of this coordinate system varies in space in respect to the highly variable size of the towers' Voronoi polygons (see Figure 1 below).

Billing of the MPN services is based on Call-Detail Records (CDR) of the Mobile Phone (MP) connections recorded by an MPN operator. The structure of the CDR for different MPN operators is similar and includes the ID of the device, ID of the antenna, ID of the cell tower that serves the connection, and the exact time interval of the connection (Ogulenko et al., 2021). Multiple records may be created during one phone talk, e.g., when the device is reconnected to another antenna while moving.

The CDR database is an excellent data source for studying human mobility. However, the possibility or, rather, the potential for privacy breaching use of these databases raises serious concerns. Today's dominant opinion is that an adversary, who obtains access to the CDR database, can identify a mobile device without knowing its number, based on a very short sequence of time-stamped field observations of the user's connection. Let us consider an adversary that follows the device user with the freely available map of cell towers and towers' VPs. This adversary can record the time of connection and, based on the VPs map, identify the location of the device in the VP-coverage-based coordinate system and the user's trajectory during the period of surveillance. The uniqueness of these trajectories is very high: several studies of the mid-2010s conclude that the probability that two or more users will repeat the same VP sequence becomes negligible starting with the 3 - 4 records only (De Montjoye et al, 2013). Erasing temporal stamp raises the identification threshold length of the sequence of 5 – 7 records, yet a very low number (Xu et al, 2017). Putting it simply, one needs to know a very short sequence of the VPs from which the device was connected to the MPN during the day, with or without timestamps, to identify the user in the CDR database, based on the cell towers used for connections, with close to 100% probability. Once the user is identified, her/his entire mobility history can be retrieved by the adversary from the CDR database, no matter if the number of the mobile phone is encrypted or not.

Being logically consistent, the VP-based locating logic does not comply with the MPN objective, which is to provide a high-quality service for every customer. In this respect, partition of space into polygons served by the cell tower makes the MPN essentially ineffective: the eventual fluctuations of the demand within the tower's service polygon will demand significant backup capacity of the tower's antennas. Real MPNs overcome this problem by instantaneously balancing antennas' loads and serving *the devices that are far beyond the cell tower's VP*. Based on the monthly distributions of distance to devices, Ogulenko et al. (2021) demonstrate that, on average, the size of the area that comprises 95% of monthly calls of tower's antennas is twice larger than the size of the corresponding Voronoi polygon and the overlap between the tower's VP and 95% service area depends on the local structure of the MPN and varies greatly. These two facts rescind the view of the VP coverage as a coordinate system and, potentially, weakens or even disables the possibility of personal identification.

The above line of arguments demands direct experimental validation. That is, we have to compare the series of the exact device's locations, during the user's everyday activities, to the location information that can be extracted from the CDR records of the user and, then, estimate whether the user was within or outside the VP of the recorded MPN cell tower at the moment of connection. In case the user is outside the cell tower VP, we have to estimate how far from that VP was the device. In this note, we report the results of such comparison based on the simultaneous locational information available from the device's CDRs on the one hand and the user's GPS location tracker on the other. To the best of our knowledge, this is the first report of this kind.

## 2. Experimental data – GPS and CRD records, and cell towers

Our analysis is based on the data of three kinds – volunteers' GPS records available from the volunteers, the volunteers' CDR records and the layer of cell towers' locations available from one of the large cellular providers in Sweden. All these data were provided for the specific research purposes to the members of our research group and with full consent of the volunteers.

### 2.1. Volunteers' GPS records

The GPS dataset investigated in this study consists of the mobility records of two volunteers, both permanent residents of the city of Uppsala, Sweden. To this end, special SIM cards were received from the provider, with USER_ID marked. These SIM cards were used for a limited period of the data collection and then disconnected from the network. Two volunteers fully consented to be observed for this purpose and kindly provided their data for this research.

During the years 2019 – 2021, the volunteers' trajectories were recorded on occasional days, when they recalled activating GPS recorders of their mobile phones. Overall, the number of tracked days for them both was 130. 68% of their GPS locations are recorded within the municipal boundary of Uppsala, almost all in the central part of the city (Figure 1a). The rest of 32% of the GPS tracks were collected when the volunteers traveled outside Uppsala (Figure 1b).

In most of the cases, the volunteers used Android mobile phones (Samsung Galaxy S4, S9, S11) and a free GPSLogger application (Mendhak et al., 2017) that automatically archived and exported their GPS tracks in GPX and CSV formats. In addition, one of them used an iPhone (models 6 and 8) with ViewRanger GPS tracking app (Augmentera, 2005). While the mobile phones' GPS tracker apps were the main source of the exact locational data, occasionally trajectories were recorded by the Canmore GT-730 GPS data logger. Below, we combine GPS tracks of all kinds. Typically, GPS tracking frequency was 1 second, while in some cases it was 1 minute. 90% of the GPS tracks were matched to the CDR records, based on the time of the MPN connections.

## 2.2. Cell towers and derived coverage of the Voronoi polygons

The locations of the MPN cell towers were used as centers of the Voronoi polygons (Figure 1).

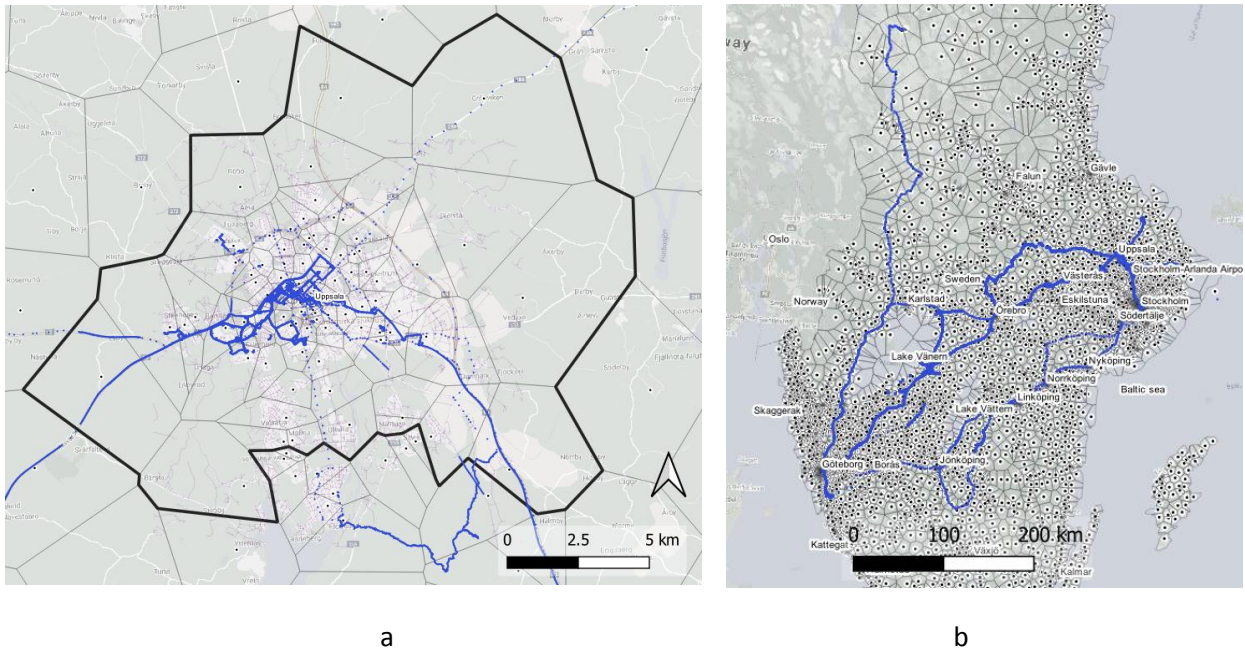

a  b

Figure 1. The overall pattern of the volunteers' GPS tracks, cell towers locations, and their VP coverage: (a) GPS tracks within the Uppsala municipality border; (b) The entire area where GPS data were collected

As can be seen in Figure 2, very roughly, the density of the cell towers is proportional to the population numbers in the VP. Each cell tower aims at serving the same number of connections and towers' service areas reflect population density. The latter results in the spatially smooth variation of the VPs' areas - the value of Moran autocorrelation coefficient between the area of the VP and the average area of the adjacent VPs is I = 0.396 (p < 0.0001).

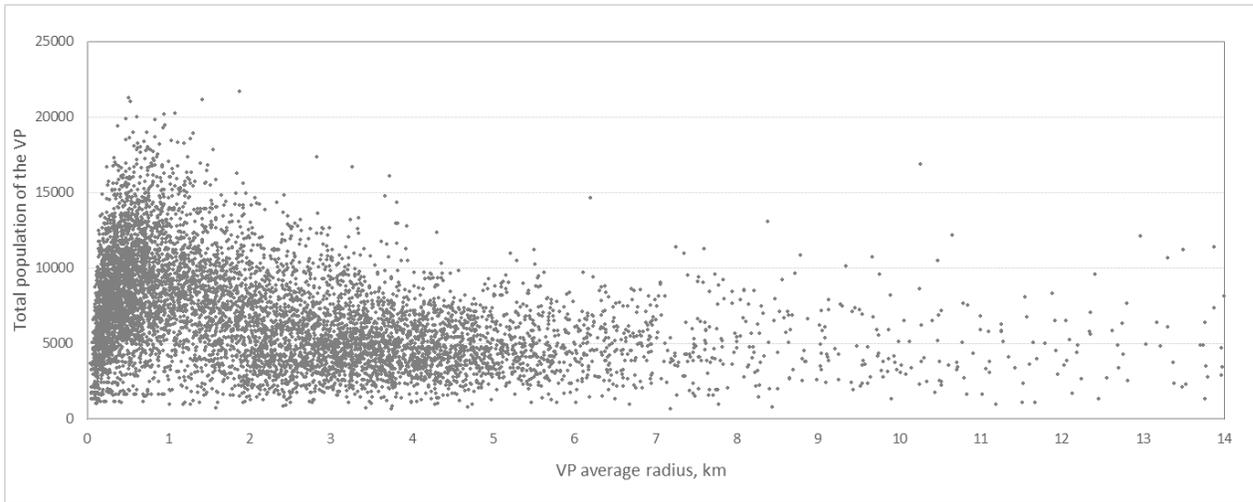

Figure 2. Total population within the cell towers' Voronoi polygons, for the polygons of the average radius of up to 10 km. The tendency remains the same for the larger polygons. Calculated based on the DeSo level population data of the Swedish statistical bureau (SCB, 2020).

Note that the average number of the adjacent neighbors of a VP is 5.88, while the median is 6, the same as of the uniform hexagonal partition (Figure 3).

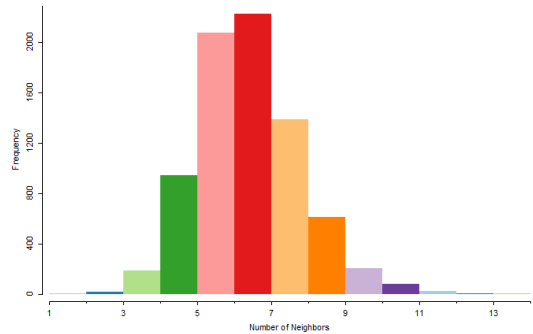

Figure 3. The distribution of the number of adjacent VPs, for each MPN VP

Figure 4 presents the size distribution of the VPs in Sweden, and the area and population covered by the VPs of a certain size.

a 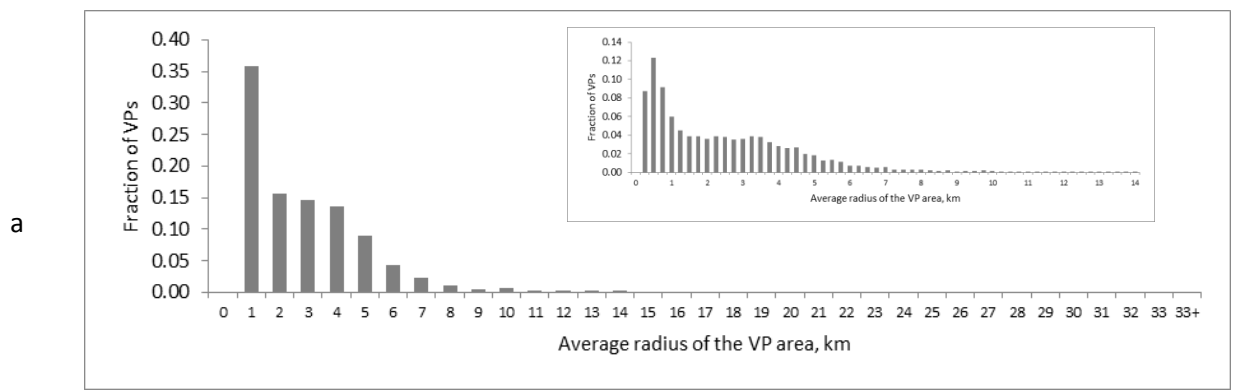

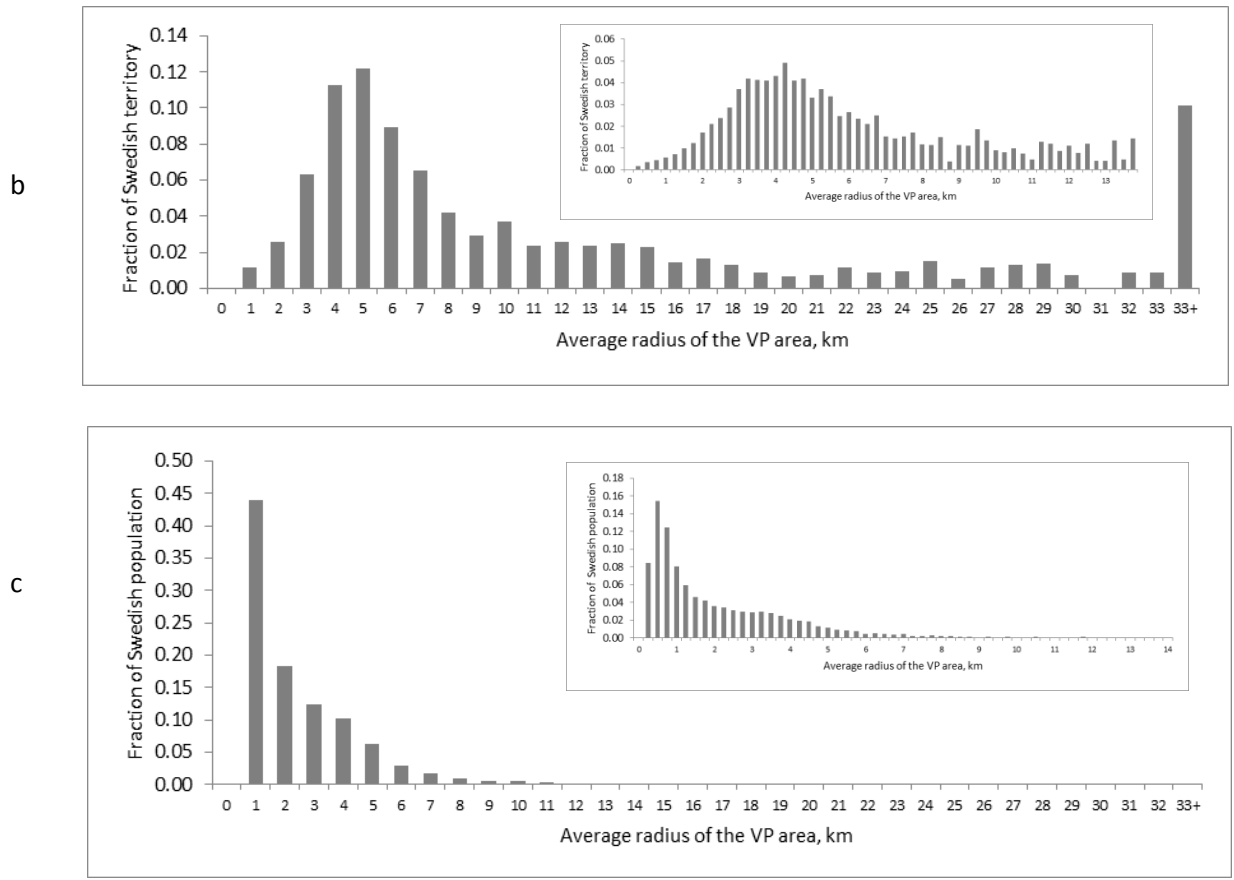

Figure 4. Size distribution of VPs in Sweden, given by the average radius of VP, whole range chart and a zoom to smaller radii (a); the fraction of the country area that is covered by the VPs of a certain size (b); and the overall population residing in the VPs of a certain size (c).

## 2.3. Call Detail Records

The data on mobile phone connections of the volunteers were retrieved from the secure MIND database established for research purposes by one of the larger GSM providers in Sweden. The MIND data server is maintained by the Uppsala University IT department for the Calista research team, whose members co-authored this study. The CDR data are aggregated over every 5 minutes at fixed time moments starting from 00:00:00. The record for the 5-minute interval contains all antennas to which the device was connected during this interval, in order of connection, but without the exact time of connection. The record consists of an anonymized USER_ID generated by the provider instead of the 'IMSI-ID', each day anew, 5-minute interval time stamp, and the ID of the MPN cells, whose antennas were used for the connection. Below, we distinguish between the cases when during the 5-min interval a user was connected to antennas of *one* or *more than one* cell tower.

Overall, the volunteers' records contain data on connections to 1124 antennas that belong to 480 cell towers. Two of these antennas, located at two different cell towers appear in the CDR records over-frequently, 487 and 336 times, respectively, while the next, by frequency, antenna appeared 97 times. These antennas were used by the volunteers when they stayed at home or work. We consider these connections as overrepresented and analyze them separately. Note that during 5 minutes of observation, the distance between the device and antenna can vary, especially when the user is driving.

## 3. Comparison of the CDR- and GPS-based device locations

We represent the distance between GPS location and the cell tower in two ways - based on the adjacency of the VPs and as the Euclidean distance. For the former representation, we consider adjacency-based ring neighborhoods of cell's VP, up to a neighborhood ring of the $6^{th}$ order (Figure 5). VP itself is considered as the ring of order 0. Below, the neighborhood of $N^{th}$ order comprises neighborhood rings of all orders between 0 and N.

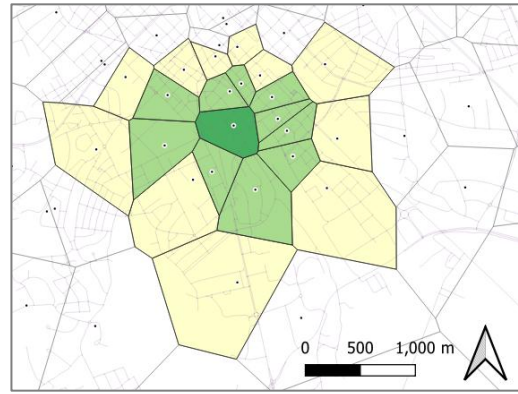

Figure 5: VP in the center of Uppsala and its neighborhood rings of orders 1 and 2.

As mentioned above, during the 5-min interval a device can be connected to a Single Cell Tower (SC) and can switch between Multiple Cell Towers (MC). For the SC case, we can determine the distance between each of the GPS locations recorded during the 5-minute interval and a cell tower. In what follows we will characterize the SC case by the closest to and the farthest from the cell tower GPS point, in terms of the absolute distance and in terms of the VP neighboring rings. Figure 6a shows this schematically.

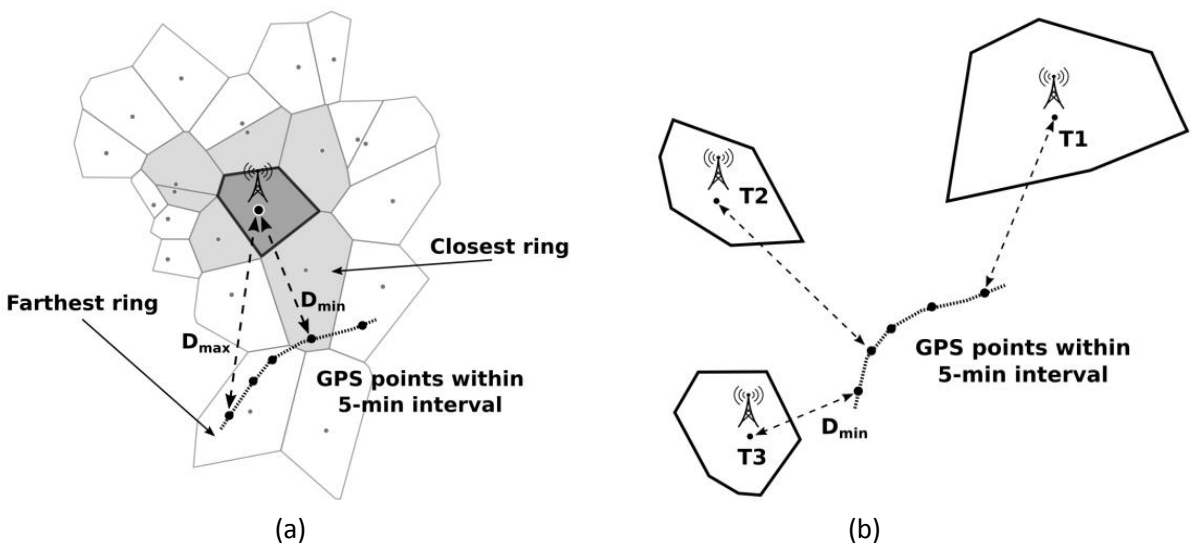

(a)          (b)

Figure 6. Estimation of the distance between the mobile device and cell tower(s) that served the device during the 5-minute interval. (a) SC case; (b) MC case.

In a device, during a 5-minute time interval, was connected to several cell towers (the MC case), we are bound to consider the minimal distance only. Typically, the MC series represent a moving device whose 5-minute GPS trajectory segment is relatively long. For example, a vehicle that drives at a speed of 72 km/h, covers 6 km during a 5-minute time interval. Since the exact time of connection to the antennas of each of the cell towers, within the 5-minute interval, is unknown, we cannot know the moment of switch between the towers. That is why in the MC case we employ the minimal distance between the cell tower and 5-minute GPS segment. This conservative estimate is always equal or lower than the actual minimal distance between the device location and the antenna (Figure 6b).

### 3.1. Analysis of connections to the single cell tower

Figure 7 presents the distribution of the minimal and maximal possible distance between the GPS location of the mobile device and the cell tower that served this connection, during a 5-minute connection. The distances are presented in VP neighborhood ring units and in kilometers. The average minimal distance is 2490 m, STD = 3480 m, while the average maximal distance is 3670 m, STD = 4830 m. In neighborhood rings, the average minimal distance is 0.73 rings, STD = 1.11 rings, while the average maximal distance is 1.12 rings, STD = 1.36 rings.

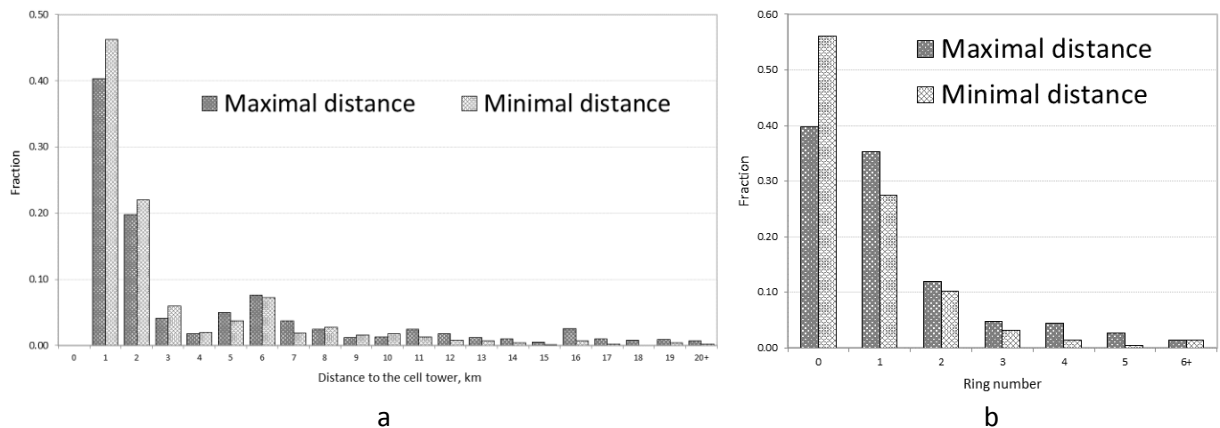

a                                              b

Figure 7. Minimal and maximal distances between the device and antennas of the cell tower that served the device during the 5-minute interval. (a) In kilometers; (b) In terms of the VP neighborhood rings

In 75% of cases, the difference in distances between the maximal and minimal distance is less than 500 m, with the rest 25% of difference values spread up to 11 km distance almost uniformly. According to the GPS data, these large differences are characteristic of driving in sparsely populated areas with low density of cell towers and large VPs. In 72% of all cases, the closest and the furthest of the 5-minute GPS observations remain inside the same VP ring, while in 21% of cases, the device moved from one neighborhood ring to another. In the rest 7% of cased, more than 1 ring were crossed during 5 minutes.

### 3.2. Analysis of connections to multiple cell towers

Figure 8 presents a histogram of the minimal distance to a cell tower for the MC connections during the 5-minute interval. The average distance to the cell tower is 3970 m, STD = 4190 m and, in the neighborhood rings, 1.23 rings, STD = 1.24 rings.

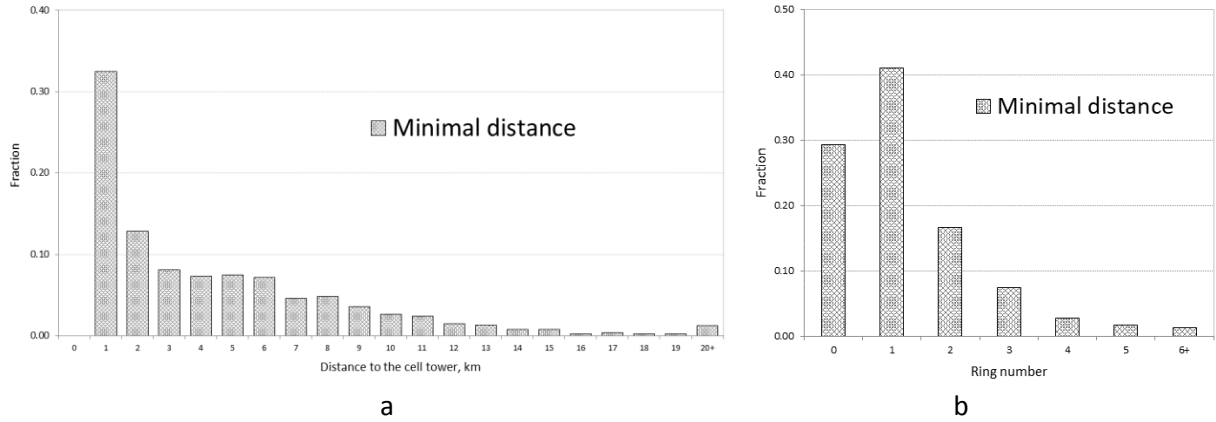

Figure 8. Histograms of the minimum distance between the device location and the cell tower in the MC case, in km (a) and in rings (b)

As can be seen, in the MC case, the share of distant locations, in kilometers and for each ring, is higher than obtained in the SC case even for the maximal distance. Possible explanation of this may be the positive dependency between the distance to the cell towers and the possibility of switching the antenna during the 5-minute time interval - a device that is far away from all cell towers the possibility to switch is higher. This explanation is confirmed by the high average minimal distance to the cell tower in the MC case, 3970 m, which is 1.5 km larger than the average minimal distance in the SC case (2490 m), and, also, larger than the average maximal distance in the SC case, which is 3670 m.

### 3.3. Combining single- and multiple cell tower connections

We continue with the histograms of the minimum distance between the 5-minute GPS segments and the cell tower that combines the MC and SC cases. The overall average minimal distance for this combined dataset is 3540 m, STD = 4050 m. In terms of the neighborhood rings, the average minimal distance is 1.08 rings, STD = 1.23 rings. To remind, the use of the minimal distance results in underestimating of the distance between the GPS segment and the cell tower.

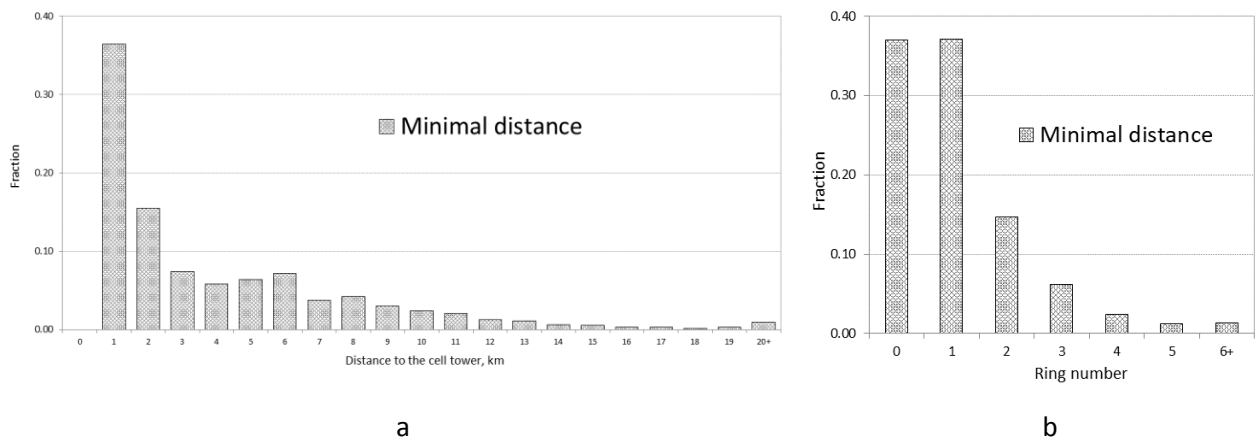

Figure 8. Histogram of the minimal distance between the location of the device and the cell tower, in kilometers (a) and in neighborhood rings (b).

In terms of Euclidean distance, the fraction of devices at a distance *d* from the cell tower is non-monotonous (Figure 8a) and an evident reason for that is the insufficient, for the high variation in the size of the cell towers' VPs, number of observation, see Figures 1 and 2.

Figure 8b presents our main result: At least 63% of all connections were performed when the device was located beyond the VP of the connection cell, with at least 37% located within the first ring, 15% within the second and 6% within the third ring of the VPs. That is, the device will be located within the neighborhood of the second order in less than 90% of all cases and within the neighborhood of the third order in less than 95% of all cases. As a rough estimate, in case of the uniform hexagonal VP partition, the number of hexagons in the $N^{th}$ ring is 6N and the size of the VP neighborhood is 1 + 3N(N+1). That is, to be sure that the neighborhood contains 90% devices that were served by a cell tower we have to consider the area that is 17 time larger that tower's VP and for the 95% percentile, the areas should be at least 37 times larger. We thus claim that the Voronoi-based view of the MPN service is wrong and must be revoked.

As expected, we found that the distance to cell tower depends on the size of the tower's VP polygon (Figure 9). The Euclidean distance to the tower evidently increases with the increase in the VP's size (Figure 9a), while in terms of rings, the distance is decreasing (Figure 9b).

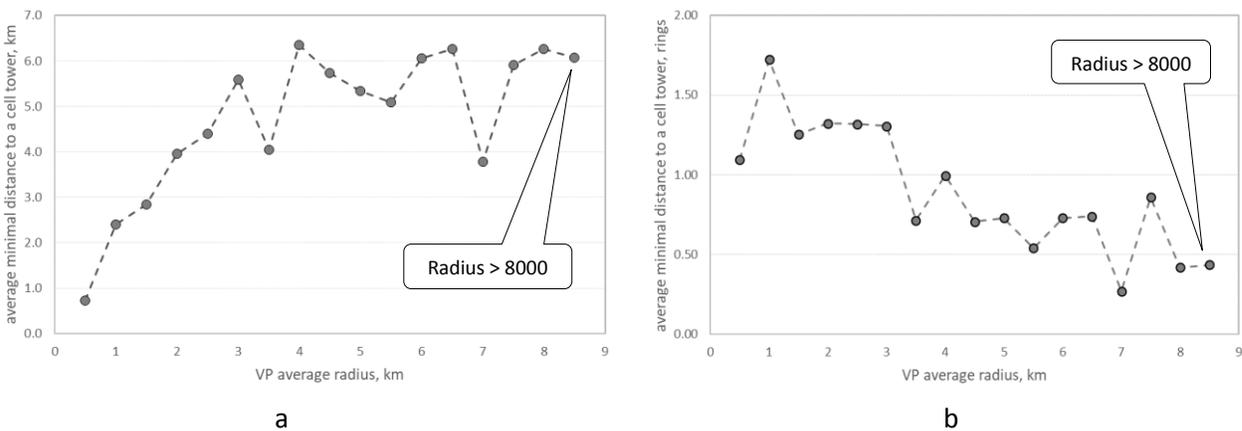

a    b

Figure 9. Average minimal distance to a cell tower, in km (a) and in rings (b), as dependent on the size of the cell tower's VP. The numbers of observations for the VPs which radius is above 8 km are low and the average over all these VPs is presented as a last point on both charts.

## 4. Analysis of volunteers' connections at frequently visited locations

As mentioned, antennas of the two specific cell towers were excluded from the above analysis due to the excessive frequency of occurrence. According to the GPS measurements, these towers are used by the volunteers when they stay at home. These connections as well as connections established by the volunteers when they were at their common workplace were studied separately and minimal distance between the GPS location and the cell towers, in rings, is presented in Table 1.

Table 1. Distribution of distances to the cell tower for the frequently visited steady locations, by VP rings

| Location | Percentage of connections performed through the VP ring | | | | | The number of connections |
| --- | --- | --- | --- | --- | --- | --- |
| | $0^{th}$ | $1^{st}$ | $2^{nd}$ | $3^{rd}$ | $4^{th}+$ | |
| Work | 50.3% | 42.5% | 2.0% | 5.2% | - | 153 |
| Home-1 | 1.7% | 7.2% | 90.1% | 0.3% | 0.6% | 362 |
| Home-2 | 1.1% | 96.3% | 0.7% | 0.5% | 1.3% | 435 |

The "Home-1" location is almost exclusively (in 90.1% of cases) served by one cell tower that belongs to the $2^{nd}$ neighboring ring of the VP that covers the location of the building. The "Home-2" location is served by the tower in the $1^{st}$ ring of its VP (96.3% of cases). The connections from the "Work" location were served by three cell towers: VP of one covers the work building, while two others are located in the $1^{st}$ neighboring ring of this VP. Figure 10 presents the distribution of the GPS locations for the sequences that were matched to the MP connections from the "work" location.

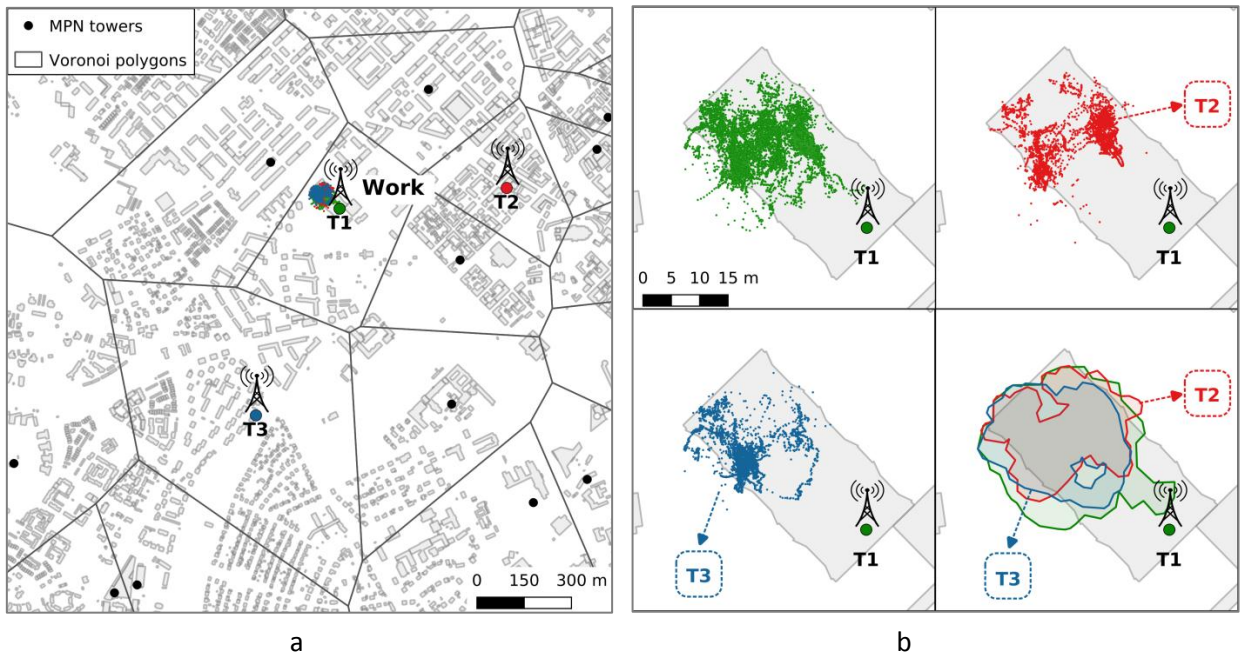

Figure 10. GPS records of the volunteers corresponding to their connections from the "Work" location. (a) Cell towers and their VPs of the orders 0 and 1 around the "Work" location, T1, T2, and T3 denote antennas that served users' communications; (b) GPS records for the time periods that match to the connections, colored by the tower of connection.

Analysis of volunteers' connections at home and at work stresses two facts: (1) Antennas of several cell towers can serve connections from the same location; (2) Connections from a certain location are not necessarily served by the closest antennas (Figure 10a); the serving antenna is assigned by the MPN based on the network state at the moment of connection.

## 5. Discussion

Basically, our results are not surprising. The MPN constantly optimizes antennas' loads in order to provide the reliable and high-quality service (Ogulenko et al., 2021). The optimization procedures deal with antennas and, typically, three antennas are mounted on a cell tower and their azimuths differ by 120°, each covering a sector of about 160° around its azimuth that is, antennas' sectors slightly overlap. The area of antenna's service has an oval shape (Zhang, 2017) and the strength of the antenna's signal within this area is non-uniform, declining to the service area border. The maximal service distance of each antenna along its azimuth is above 32 km (Zhang, 2017), when not intentionally limited by the operator. Essentially overlap of antennas' service areas limit the use of towers' Voronoi polygons to the logical construction used for explaining the principles on the MPN pattern and functioning. Wide overlapping areas of antennas' service provide necessary flexibility in choosing the location of cell towers that serve the stands for the antennas. Towers' location is dictated and constrained by many factors like terrain and buildings' 3D patterns, distribution of the resident and visiting population, planning and regulation, temporal activity of the potential customers and infrastructure-related constraints.

### 5.1. How robust are our results?

We cannot guarantee that the volunteers' data as well as local (Uppsala and nearby region) and global parameters of the MPN that we investigate are representative and that the mobility data of additional volunteers would not change the results qualitatively. Yet we believe that a long period of observation during which the volunteers covered significant part of Uppsala city makes our results representative and robust. The papers we are aware of, focus on estimating the uncertainty of the CDR-based estimates of devices single location or trajectory based on the comparison between the GPS- and CDR-based estimates of the device's location. All these papers confirm our results pointing to essential deviation between the GPS- and CDR-based estimates and suggest the methods of reducing this deviation by averaging several CDR-based observations or projecting CDR-based location on road links. Zandbergen (2009) was the first who estimated the difference between the GPS location of the 3G iPhone3 and location of the cell tower that served the connection. The estimates of the difference vary between 0.5 and 2.7 km with the median difference of 600 m. Recent and more extensive study of Pappalardo (2021) reports larger differences between the CDR- and GPS-based estimates of the home location, which vary, depending on the location between 1 and 5 km and just as observed in our study, the antenna that serves the connection is not necessarily the closest one. The average difference between CDR- and GPS trajectories varies between several hundred meters to 1-2 km (Horn et al 2014; Lind et al, 2017), while the estimates of the total traversed distance are closer, and the difference does not exceed 20% (Mohamed et al, 2017).

The closest to ours, the study by Ulm et al (2015) compares GPS and cell tower location data for 250 volunteers, available from the OpenCellID project (https://opencellid.org/) that collects data on cell tower connections together with the GPS positions. Different from our approach, the position of the cell tower was not known a priory and is estimated as a center point of volunteers' connections (Clarkson et al, 1996). The distribution of the differences between the CDR- and GPS-based estimates of the device location is qualitatively similar to that obtained in our research: skewed to the left, with a long tail.

However, its mode is essentially lower, about 200-300 m and the tail is essentially shorter, about 5 km (Ulm et al, 2015). The reason of this underestimate can be unknown exact position of the cell tower.

## 5.2. Consequences for location privacy

Contemporary privacy legislation and regulation is rooted in the privacy framework established by OECD in 1980[1], often referred to as the "fair information practices". According to Cavoukian (1998), data mining activity may come into collision with five basic principles of (1) Data Quality; (2) Purpose Specification; (3) Use Limitation; (4) Openness; (5) Individual Participation. In regards to the Openness principle, Cavoukian stresses that "Data mining … is invisible. Data mining technology makes it possible to analyze huge amounts of information about individuals — their buying habits, preferences, and whereabouts … without their knowledge or consent". By design, data mining encourages the information collected for one purpose to be used for another one, strongly contradicting the Use Limitation principle (Cavoukian, 1998). Authors of the "fair information practices" could hardly imagine the amount of data collected and the level of penetration of the data mining technologies into human activities nowadays. The collision between data mining and privacy protection led to legal challenges and inspired privacy enhancing technologies. Almost every new approach proposed as a general solution was soon adapted to privacy protection in location data (Calabrese et al., 2014; Chatzikokolakis et al., 2017; Wang et al., 2018).

A decade ago it became clear that the simple obfuscation methods of spatial cloaking, mix-zones, path confusion, and dummy trajectories (Chow & Mokbel, 2011) are not enough to guarantee the acceptable level of location privacy. Concurrently, the concept of *k*-anonymity (Samarati & Sweeney, 1998) was used to measure microscopic location-privacy (Gruteser & Grunwald, 2003) and gave birth to *l*-diversity that deals with the usage of background knowledge for attack on privacy. (Machanavajjhala et al., 2007). Yet, *k*-anonymity was evaluated as "a tattered cloak for protecting location privacy" (Shokri et. al., 2010).

Another direction - differential privacy (Dwork et al., 2006) aims at protecting individual's data stored in the database while making it possible to publish the aggregate information. The idea is to add a controlled noise to the query outcome. Since its formal introduction, differential privacy became ubiquitous and in 2020, Google declared[2] differential privacy as its main tool for personal information processing. The natural application of differential privacy to location data was in the form of the differential perturbations technique and Dewri (2013) proved that these perturbations can be generated in a way that does not bear any linkage to a single individual. Further on, Andrés et al. (2013) proposed the notion of geo-indistinguishability: the user's real location is indistinguishable from any other location within a radius *r* - the closer are the locations the harder it is for the attacker to determine the real one. Salas et al. (2020) demonstrated that sanitizing the data can be used to preserve the origin-destination matrices while producing sanitized trajectories.

Aforementioned theoretical concepts in conjunction with the avalanche of location data produced by mobile phones and GPS devices originated a vast number of applied studies (Calabrese et al., 2014; Chen et al., 2016; Wang et al., 2018; Huang et al., 2019). Important from our perspective, is that most of them share the common, but rarely stated explicitly assumption: the data supplied by the GPS equipment and mobile phone operators contain the *exact* and *certain* information about user's whereabouts. Almost certainly, it is true for the GPS data, where, privacy risks can be assessed in a precise and effective way (Pellungrini et al., 2017).

---

[1] https://www.oecd.org/sti/ieconomy/oecdguidelinesontheprotectionofprivacyandtransborderflowsofpersonaldata.htm
[2] https://blog.google/technology/safety-security/our-work-keep-you-safe-and-control-your-privacy/

The accuracy and certainty of mobile phone data are much more questionable (Ricciato et al., 2017; Bachir et al., 2019; Tennekes et al., 2020; Ogulenko et al., 2021). Yet, many studies consider locations of cell towers recorded in the database as a proxy for the device location. In many studies of individual identification through mobility, the individual's trajectory is considered formally, as a time-annotated sequence of abstract elements denoting location. Xu et al. (2017) state this explicitly "by looking up the locations of these base stations, we are able to obtain trajectories of mobile users, which serve as the ground truth in the investigation of aggregated mobility data's privacy leakage." Zang & Bolot (2011) represent a location using the combination <switch ID>-<cell ID>-<antenna sector ID> and perform a thorough analysis of the trajectories' similarity and individual's anonymity, setting aside the relationship between this formal "label" and the real location.

In reality, the very function of MPN implies essential overlap between the service areas of antennas and uncertainty in a choice of the antenna for establishing connection. In these circumstances, the certainty of locating a device is significantly lower than stated according to the Voronoi-based view (Ogulenko et al., 2021). Our study suggests that one has to consider an area that is essentially wider than a cell tower VP as the possible area of the device location. The VP's neighborhoods of the $2^{nd}$ or $3^{d}$ order may be the candidates for these areas. Consequently, the areas of possible location for two devices, whose connections were registered by two close cell towers, may essentially overlap and, thus, it is impossible for the adversary to determine which served which. The account for this uncertainty demands a new approach to estimating location privacy, urban mobility and, especially the Origin-Destination matrices based on the CDR records. The latter is the topic of our current research.